\def\maketitle{\par
 \begingroup
 \def\thefootnote{\fnsymbol{footnote}}
 \def\@makefnmark{\mbox{$^\@thefnmark$}}
 \@maketitle
 \@thanks
 \endgroup
 \setcounter{footnote}{0}
 \let\maketitle\relax
 \let\@maketitle\relax
 \gdef\@thanks{}\gdef\@author{}\gdef\@title{}\let\thanks\relax}
\def\@maketitle{\vspace*{0.9cm}
{\hsize\textwidth
 \linewidth\hsize \centering
 {\normalsize \bf \@title \par} \vskip 1.0cm  {\normalsize  \@author \par}}}

\def\thefootnote{\mbox{\noindent$\fnsymbol{footnote}$}}
    \long\def\@makefntext#1{\noindent$^{\@thefnmark}$#1}

\def\section{\@startsection{section}{1}{\z@}{1.5ex plus 0.5ex minus
   1.2ex}{1.3ex plus .1ex}{\normalsize\bf}}
\def\subsection{\@startsection{subsection}{2}{\z@}{1.5ex plus 0.5ex minus
    1.2ex}{1.3ex plus .1ex}{\normalsize\em}}

\def\@sect#1#2#3#4#5#6[#7]#8{\ifnum #2>\c@secnumdepth
     \def\@svsec{}\else
     \refstepcounter{#1}\edef\@svsec{\ifnum #2=1 \@sectname\fi
        \csname the#1\endcsname.\hskip 1em }\fi
     \@tempskipa #5\relax
      \ifdim \@tempskipa>\z@
        \begingroup #6\relax
          \@hangfrom{\hskip #3\relax\@svsec}{\interlinepenalty \@M #8\par}
        \endgroup
       \csname #1mark\endcsname{#7}\addcontentsline
         {toc}{#1}{\ifnum #2>\c@secnumdepth \else
                      \protect\numberline{\csname the#1\endcsname}\fi
                    #7}\else
        \def\@svsechd{#6\hskip #3\@svsec #8\csname #1mark\endcsname
                      {#7}\addcontentsline
                           {toc}{#1}{\ifnum #2>\c@secnumdepth \else
                             \protect\numberline{\csname the#1\endcsname}\fi
                       #7}}\fi
     \@xsect{#5}}

\def\@sectname{}

\def\thebibliography#1{\section*{{{\normalsize
\bf References }
\rule{0pt}{0pt}}\@mkboth
  {REFERENCES}{REFERENCES}}\list
  {{\arabic{enumi}.}}{\settowidth\labelwidth{{#1}}%
    \leftmargin\labelwidth  \frenchspacing
    \advance\leftmargin\labelsep
    \itemsep=-0.2cm
    \usecounter{enumi}}
    \def\newblock{\hskip .11em plus .33em minus -.07em}
    \sloppy
    \sfcode`\.=1000\relax}

\def\@cite#1#2{\unskip\nobreak\relax
    \def\@tempa{$\m@th^{\hbox{\the\scriptfont0 #1}}$}%
    \futurelet\@tempc\@citexx}
\def\@citexx{\ifx.\@tempc\let\@tempd=\@citepunct\else
    \ifx,\@tempc\let\@tempd=\@citepunct\else
    \let\@tempd=\@tempa\fi\fi\@tempd}
\def\@citepunct{\@tempc\edef\@sf{\spacefactor=\the\spacefactor\relax}\@tempa
    \@sf\@gobble}

\def\citenum#1{{\def\@cite##1##2{##1}\cite{#1}}}
\def\citea#1{\@cite{#1}{}}

\newcount\@tempcntc
\def\@citex[#1]#2{\if@filesw\immediate\write\@auxout{\string\citation{#2}}\fi
  \@tempcnta\z@\@tempcntb\m@ne\def\@citea{}\@cite{\@for\@citeb:=#2\do
    {\@ifundefined
       {b@\@citeb}{\@citeo\@tempcntb\m@ne\@citea\def\@citea{,}{\bf ?}\@warning
       {Citation `\@citeb' on page \thepage \space undefined}}%
    {\setbox\z@\hbox{\global\@tempcntc0\csname b@\@citeb\endcsname\relax}%
     \ifnum\@tempcntc=\z@ \@citeo\@tempcntb\m@ne
       \@citea\def\@citea{,}\hbox{\csname b@\@citeb\endcsname}%
     \else
      \advance\@tempcntb\@ne
      \ifnum\@tempcntb=\@tempcntc
      \else\advance\@tempcntb\m@ne\@citeo
      \@tempcnta\@tempcntc\@tempcntb\@tempcntc\fi\fi}}\@citeo}{#1}}
\def\@citeo{\ifnum\@tempcnta>\@tempcntb\else\@citea\def\@citea{,}%
  \ifnum\@tempcnta=\@tempcntb\the\@tempcnta\else
   {\advance\@tempcnta\@ne\ifnum\@tempcnta=\@tempcntb \else \def\@citea{--}\fi
    \advance\@tempcnta\m@ne\the\@tempcnta\@citea\the\@tempcntb}\fi\fi}

\def\abstract{\if@twocolumn
\section*{Abstract}         
\else \small
\begin{center}
{ABSTRACT\vspace{-.5em}\vspace{0pt}}
\end{center}
\quotation
\fi}
\def\endabstract{\if@twocolumn\else\endquotation\fi}

\def\fnum@figure{Fig. \thefigure}

\long\def\@makecaption#1#2{
   \vskip 10pt
   \setbox\@tempboxa\hbox{\small #1. #2}
   \ifdim \wd\@tempboxa >\hsize    
      \small #1. #2\par            
   \else                           
      \hbox to\hsize{\hfil\box\@tempboxa\hfil}
   \fi}

\documentstyle[12pt,titlepage]{article}
\begin{document}
\pagestyle{empty}
\newcommand{\lapproxeq}{\lower
.7ex\hbox{$\;\stackrel{\textstyle <}{\sim}\;$}}
\newcommand{\skips}{\vspace{0.2cm}}
\newcommand{\skipm}{\vspace{0.4cm}}
\baselineskip=0.212in

\begin{flushleft}
\large
{SAGA-HE-48  \hfill September 30, 1993}  \\
\end{flushleft}

\vspace{2.5cm}

\noindent
\Large{{\bf 1)~~  Spin and flavor sum rules based on a parton model}} \\

\vspace{0.5cm}

\noindent
\Large{{\bf 2)~~  SU(2) flavor breaking in antiquark distributions}} \\

\vspace{1.8cm}

\begin{center}

\Large
{S. Kumano}         \\

\vspace{1.3cm}

\Large
{Department of Physics} \\

\vspace{0.2cm}

\Large{Saga University}    \\

\vspace{0.2cm}

\Large
{Saga 840, Japan}         \\

\vspace{2.5cm}

\Large
{Invited and Contributed Talks} \\

{at the XIII International
 Conference on Particles and Nuclei} \\

{Perugia, Italy, June 28 $-$ July 2, 1993} \\

\end{center}

\end{document}